\documentstyle[twoside,fleqn,espcrc2,epsf,rotate]{article}

\newcommand{\nn}{\nonumber}
\newcommand{\eq}[1]{(\ref{#1})}
\newcommand{\fig}[1]{fig.\ \ref{#1}}
\newcommand{\tab}[1]{tab.\ \ref{#1}}
\newcommand{\MS}{{\rm MS}}
\newcommand{\MSbar}{\overline{\MS}}
\newcommand{\GF}{{\rm G}_{\rm F}}
\newcommand{\MW}{M_{\rm W}}

\renewcommand{\floatsep}{0pt}
\renewcommand{\textfloatsep}{5pt}
\renewcommand{\intextsep}{10pt}

\title{
	\hfill {\normalsize TUM-T31-72/94}\\[-2truemm]
	\hfill {\normalsize hep-ph/9408398}\\[-2truemm]
	\hfill {\normalsize August 94}\\
	Next-to-leading Order short distance QCD corrections to the
	effective $\Delta S = 2$ Hamiltonian, Implications for the
	${\rm K}_{\rm L}$--${\rm K}_{\rm S}$ mass difference
	\thanks{
		Invited talk presented at the conference ``QCD'94'',
		Montpellier, France, July 7 - 13, 1994.
		To appear in the proceedings.
		}
	}
\author{
	Stefan Herrlich
	\address{
		Physik-Department T31, Technische Universit\"at
		M\"unchen, D-85747 Garching, Germany
		}
	\thanks{
		E-Mail: herrl@feynman.t30.physik.tu-muenchen.de
		}
	}

\begin{document}

\begin{abstract}
We report on the results of a calculation of next-to leading order
short distance QCD corrections to the coefficient $\eta_1$ of the
effective $\Delta S = 2$ Lagrangian in the standard model and discuss
the uncertainties inherent in such a calculation.
As a phenomenological application we comment on the contributions of
short distance physics to the ${\rm K}_{\rm L}$--${\rm K}_{\rm S}$
mass difference.
\end{abstract}

\maketitle

\section{Introduction}
This report is based on research work done in collaboration with Ulrich
Nierste \cite{HerrlichNierste:94}.

The prediction of observables in the ${\rm K^{0}}$--$\overline{\rm
K^{0}}$ system forces one to calculate an effective low-energy $\Delta
S = 2$ hamiltonian, which is difficult because of the necessary
inclusion of strong interaction effects.
Applying Wilson's operator product expansion factorizes the Feynman
amplitude into a long-distance part, to be evaluated by
non-perturbative methods, and a short-distance part, which can be
calculated in renormalization group improved perturbation theory.

\begin{figure}[htb]
	\epsfysize=2truecm
	\epsffile{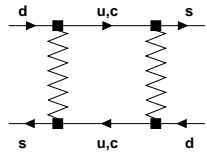}
\vspace{-25truept}
\caption{The box diagrams contributing to $H$ in lowest order.
	The zigzag lines represent ${\rm W}$-bosons or fictitious
	Higgs particles.
	}
\label{fig:box}
\end{figure}

The short distance part of the $\Delta S = 2$ Feynman amplitude has
first been calculated to leading order (LO) by
Vainstein et al.\ \cite{VainsteinZakharovNovikovShifman:76},
Vysotskij \cite{Vysotskij:80} and
Gilman and Wise \cite{GilmanWise:83}.
These determinations leave certain questions unanswered which we like
to summarize below.
To be specific, consider the effective $\Delta S = 2$ hamiltonian
\begin{eqnarray}
	H
	&=&
	\frac{\GF^2}{16\pi^2}\MW^2
	\biggl[
		\lambda_{\rm c}^2 \eta_1
			S\!\left(\textstyle\frac{m_{\rm c}^2}{\MW^2}\right)
		+ \lambda_{\rm t}^2 \eta_2 \times
\nn \\
	& &
			S\!\left(\textstyle\frac{m_{\rm t}^2}{\MW^2}\right)
		+ 2 \lambda_{\rm c} \lambda_{\rm t} \eta_3
			S\!\left(\textstyle\frac{m_{\rm c}^2}{\MW^2},
				\frac{m_{\rm t}^2}{\MW^2}\right)
	\biggr]
\nn \\
	& &
	\times b\!\left(\mu\right) Q_{\rm LL}\!\left(\mu\right)
\nn \\
	&=&
	H^{\rm c} + H^{\rm t} + H^{\rm ct}
\label{Heff}
\end{eqnarray}
with $\GF$ denoting Fermi's constant,
$\lambda_{j} = V_{j{\rm d}}^{*} V_{j{\rm s}}$, $j={\rm c,t}$ the
relevant combination of CKM factors and $Q_{\rm LL}$ the local
four-quark operator
\begin{eqnarray}
	Q_{\rm LL}
	&=&
	\left[\bar{\rm s} \gamma_{\mu} \left(1-\gamma_{5}\right)
		{\rm d} \right]
	\left[\bar{\rm s} \gamma^{\mu} \left(1-\gamma_{5}\right)
		{\rm d} \right]
\label{QLL}
\end{eqnarray}
In \eq{Heff} the GIM mechanism $\lambda_{\rm u}+\lambda_{\rm
c}+\lambda_{\rm t} = 0$ has been used to eliminate $\lambda_{\rm u}$.
The Inami-Lim functions $S$ are obtained in the evaluation of the
famous box diagrams depicted in \fig{fig:box}.
The $\eta_{i}$ parametrize the short distance QCD corrections with
their explicit dependence on the renormalization scale $\mu$ factored
out in the function $b\!\left(\mu\right)$.
In absence of QCD corrections $\eta_{i} b\!\left(\mu\right) = 1$.

The LO calculation leaves the following questions unanswered
\begin{list}{$\bullet$}{
	\setlength{\labelwidth}{10pt}
	\setlength{\labelsep}{2pt}
	\setlength{\leftmargin}{6pt}
	\setlength{\itemsep}{-3pt}
	\setlength{\topsep}{0pt}
	\setlength{\parsep}{3pt}
	}
\item The running charm quark mass $m_{\rm c}$ enters \eq{Heff} at the
scale $\mu_{\rm c} \approx m_{\rm c}$, where the dynamic charm quark
is removed from the theory.
The LO result depends strongly on the choice of $\mu_{\rm c}$.
A similar statement applies to the scale $\mu_{\rm W} \approx \MW$,
where the ${\rm W}$-boson and the top quark are integrated out.
\item The precise definition of the QCD scale parameter $\Lambda_{\rm
QCD}$ requires at least a next-to-leading order (NLO) calculation.
\item Subleading terms may contribute sizeable, e.g.\ the LO
hamiltonian reproduces only about 60\% of the observed ${\rm K_{L}}$--
${\rm K_{S}}$ mass difference.
\end{list}
Prior to our work the only part of \eq{Heff} calculated to NLO was
$H^{\rm t}$ containing $\eta_2$ \cite{BurasJaminWeisz:90}.
This report deals with $H^{\rm c}$ and the coefficient $\eta_1$.

\section{The NLO calculation}
All calculations are carried out in the $\MSbar$ scheme using an
arbitrary $R_{\xi}$ gauge for the gluon propagator and
't\ Hooft--Feynman gauge for the ${\rm W}$-propagator.
Inspired by refs.\ \cite{BurasWeisz:90,BurasJaminWeisz:90}
we use an anticommuting $\gamma_{5}$ (NDR scheme).
Infrared singularities get regulated by small quark masses.
We only calculate the lowest nonvanishing order in $m_{\rm c}/\MW$,
therefore setting $S\!\left(m_{\rm c}^2/\MW^2\right)=m_{\rm c}^2/\MW^2$.
This turns out to be necessary, if we only want to keep operators of
the lowest twist.

\begin{figure}
	\epsfysize=1.5truecm
	\epsffile{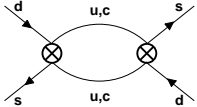}
\vspace{-25truept}
\caption{
	The lowest order diagram contributing to the effective five
	and four flavour theories between the scales $\mu_{\rm W}$ and
	$\mu_{\rm c}$.
	The crosses denote the insertion of local $\Delta S=1$
	operators.
	}
\label{fig:eye}
\end{figure}

The calculation is performed in the standard renormalization group
technique.
The matching of different theories at matching scales
$\mu_{\rm W}$, $\mu_{\rm b}$ and $\mu_{\rm c}$
requires the evaluation of the diagram in \fig{fig:eye} and diagrams
derived from this by dressing it with one gluon in all possible
ways.
No local $\Delta S=2$ operator appears in the effective five and four
quark theory, because the diagrams mentioned above are finite due
to the GIM mechanism.
Such an operator, it is the $Q_{\rm LL}$ of \eq{QLL}, first arises
after moving to an effective three flavour theory by removing the
${\rm c}$-quark.
For details, we refer the reader to \cite{HerrlichNierste:94}.

To check our calculation, we performed various consistency tests.
\begin{list}{$\bullet$}{
	\setlength{\labelwidth}{10pt}
	\setlength{\labelsep}{2pt}
	\setlength{\leftmargin}{6pt}
	\setlength{\itemsep}{-3pt}
	\setlength{\topsep}{0pt}
	\setlength{\parsep}{3pt}
	}
\item
The Wilson coefficient functions turn out to be independent of the
gluon gauge parameter $\xi$ and the small quark masses used as a
regulator for the infrared singularities.
\item
Setting $\mu_{\rm W}=\mu_{\rm b}=\mu_{\rm c}$, we recover the result
obtained in naive perturbation theory up to the first order in the
strong coupling constant $\alpha_{\rm s}$.
\item
The dependence of the final result on the matching scales $\mu_{\rm
W}$, $\mu_{\rm b}$, $\mu_{\rm c}$ vanishes up to first order in
$\alpha_{\rm s}$.
\end{list}

\section{Numerical results}
\begin{figure}[htb]
	\rotate[r]{
	\epsfysize=7truecm
	\epsffile{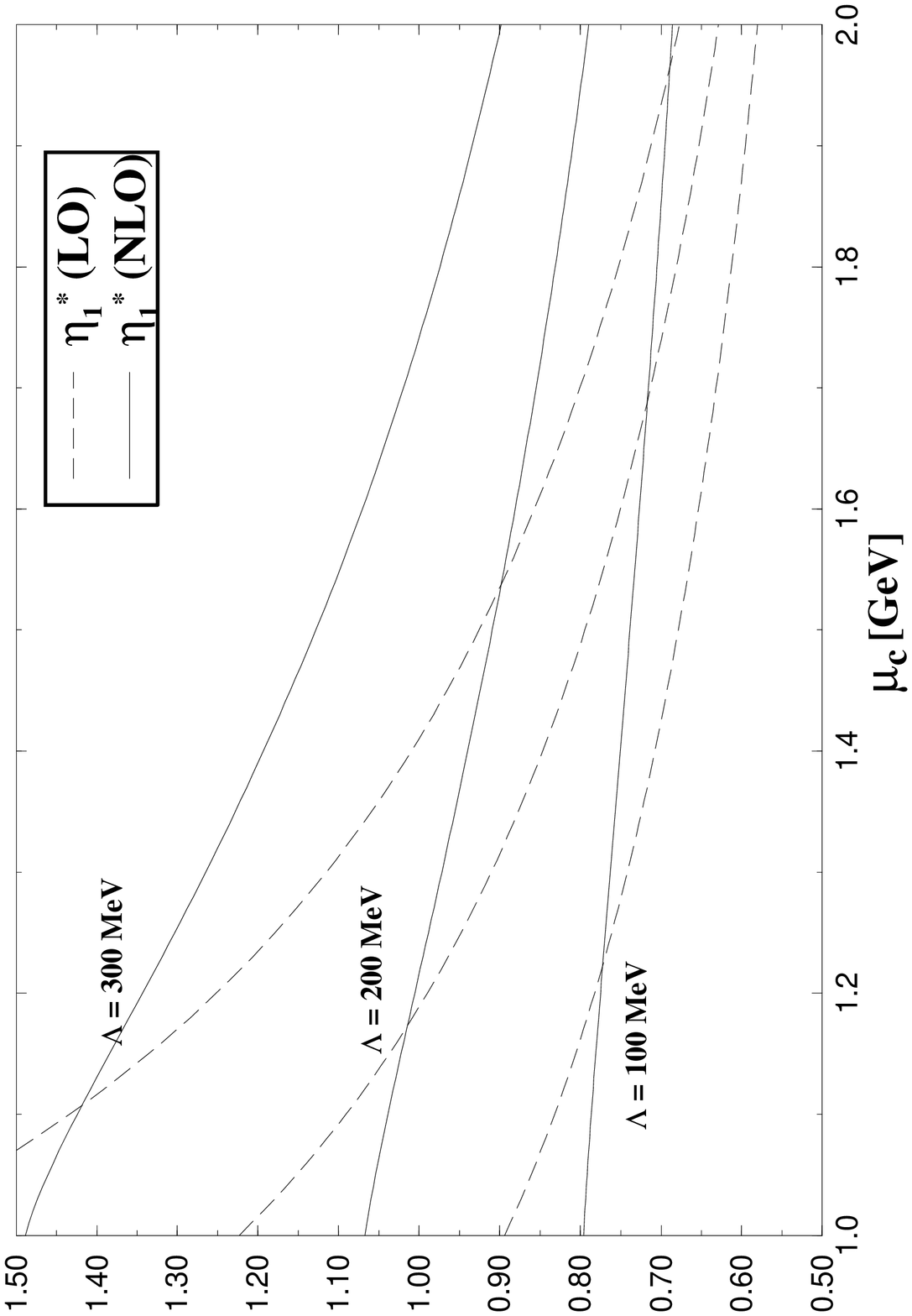}
	}
\vspace{-25truept}
\caption{
	The variation of $\eta_1^{\star}$ with the matching scale
	$\mu_{\rm c}$ for different values of $\Lambda_{\overline{\rm
	MS}}=\Lambda_{4}$ and $\mu_{\rm c}$.
	$\mu_{\rm W}=\MW=80{\rm GeV}$, $\mu_{\rm b}=4.8{\rm GeV}$ and
	$m_{\rm c}^{\star}=1.4{\rm GeV}$ have been kept fixed.
	}
\label{fig:scaledep}
\end{figure}

In the analytical calculation the running charm quark mass
renormalized at the scale $\mu_{\rm c}$, where it gets incorporated
into the Wilson coefficient function, enters quite naturally.
For the numerical analysis and the discussion of renormalization scale
dependence it is better to use
$m_{\rm c}^{\star} = m_{\rm c}\!\left(m_{\rm c}\right)$.
We therefore define $\eta_1^{\star}$ by
\begin{eqnarray}
	\eta_{1}\!\left(\mu_{\rm W},\mu_{\rm c}\right)
	m_{\rm c}^2\!\left(\mu_{\rm c}\right)
	&=&
	\eta_{1}^{\star}\!\left(\mu_{\rm W},\mu_{\rm c}\right)
	\left(m_{\rm c}^{\star}\right)^2
	.
\label{EtaStar}
\end{eqnarray}
In \tab{tab:eta} we listed $\eta_1^{\star}$ for different values of
the QCD scale parameter in the effective four flavour theory
$\Lambda_{\overline{\rm MS}}=\Lambda_{4}$ and $m_{\rm c}^{\star}$.
\begin{table*}[t]
\caption{
	$\eta_1^{\star}$ for different values of $m_{\rm c}^{\star}$
	and $\Lambda_{\overline{\rm MS}}=\Lambda_{4}$ (both given in
	GeV).
	The matching scales have been taken to be
	$\mu_{\rm W}=\MW=80{\rm GeV}$, $\mu_{\rm b}=4.8{\rm GeV}$ and
	$\mu_{\rm c}=m_{\rm c}^{\star}=m_{\rm c}\!\left(m_{\rm
	c}\right)$.
	}
%
%
\begin{tabular*}{\textwidth}{@{}c@{\extracolsep{\fill}}cccccccccc}
\hline
$\Lambda_{\overline {\rm MS}}$
 & \multicolumn{2}{c}{0.150}
 & \multicolumn{2}{c}{0.200}
 & \multicolumn{2}{c}{0.250}
 & \multicolumn{2}{c}{0.300}
 & \multicolumn{2}{c}{0.350}
\\
\cline{2-3} \cline{4-5} \cline{6-7} \cline{8-9} \cline{10-11}
$m_c^{\star}$ & LO & NLO & LO & NLO & LO & NLO & LO & NLO & LO & NLO\\
\hline
1.25 &  0.809 &  0.885 &  0.895 &  1.007 &  0.989 &  1.154 &  1.096 &  1.334 &
1.216 &  1.562 \\
1.30 &  0.797 &  0.868 &  0.877 &  0.982 &  0.965 &  1.117 &  1.064 &  1.281 &
1.175 &  1.485 \\
1.35 &  0.786 &  0.854 &  0.861 &  0.960 &  0.944 &  1.085 &  1.035 &  1.235 &
1.138 &  1.419 \\
1.40 &  0.775 &  0.840 &  0.847 &  0.940 &  0.924 &  1.056 &  1.010 &  1.194 &
1.105 &  1.361 \\
1.45 &  0.766 &  0.828 &  0.834 &  0.922 &  0.907 &  1.030 &  0.987 &  1.157 &
1.075 &  1.310 \\
1.50 &  0.757 &  0.817 &  0.822 &  0.905 &  0.890 &  1.006 &  0.966 &  1.125 &
1.048 &  1.265 \\
1.55 &  0.749 &  0.806 &  0.810 &  0.890 &  0.876 &  0.985 &  0.946 &  1.095 &
1.024 &  1.225 \\
\hline
\end{tabular*}
\label{tab:eta}
\end{table*}
\begin{table*}[t]
\caption{
	$\Delta m_{\rm K}^{\rm SD,c}$ in NLO in units of $10^{-15}{\rm
	GeV}$ for different values of $\Lambda_{\overline{\rm
	MS}}=\Lambda_{4}$ and $\mu_{\rm c}=m_{\rm c}^{\star}$ (both
	given in GeV).
	$B_{\rm K} = 0.7$
	}
%
%
%
\begin{tabular*}{\textwidth}{@{}c@{\extracolsep{\fill}}cccccccccc}
\hline
$\Lambda_{\overline {\rm MS}}$
 & \multicolumn{2}{c}{0.150}
 & \multicolumn{2}{c}{0.200}
 & \multicolumn{2}{c}{0.250}
 & \multicolumn{2}{c}{0.300}
 & \multicolumn{2}{c}{0.350}
\\
\cline{2-3} \cline{4-5} \cline{6-7} \cline{8-9} \cline{10-11}
$m_c^{\star}$ & $\Delta m$ & $\frac{\Delta m}{\Delta m_{\rm EXP}}$ & $\Delta m$
& $\frac{\Delta m}{\Delta m_{\rm EXP}}$ & $\Delta m$ & $\frac{\Delta m}{\Delta
m_{\rm EXP}}$ & $\Delta m$ & $\frac{\Delta m}{\Delta m_{\rm EXP}}$ & $\Delta m$
& $\frac{\Delta m}{\Delta m_{\rm EXP}}$\\
\hline
1.25 &  1.327 &  0.377 &  1.510 &  0.429 &  1.730 &  0.491 &  2.000 &  0.568 &
2.342 &  0.665 \\
1.30 &  1.408 &  0.400 &  1.593 &  0.452 &  1.812 &  0.514 &  2.078 &  0.590 &
2.409 &  0.684 \\
1.35 &  1.493 &  0.424 &  1.679 &  0.477 &  1.897 &  0.539 &  2.159 &  0.613 &
2.481 &  0.705 \\
1.40 &  1.580 &  0.449 &  1.768 &  0.502 &  1.986 &  0.564 &  2.245 &  0.637 &
2.560 &  0.727 \\
1.45 &  1.670 &  0.474 &  1.860 &  0.528 &  2.078 &  0.590 &  2.335 &  0.663 &
2.643 &  0.751 \\
1.50 &  1.763 &  0.501 &  1.955 &  0.555 &  2.173 &  0.617 &  2.428 &  0.689 &
2.731 &  0.776 \\
1.55 &  1.859 &  0.528 &  2.052 &  0.583 &  2.271 &  0.645 &  2.525 &  0.717 &
2.824 &  0.802 \\
\hline
\end{tabular*}
\label{tab:deltaM}
\end{table*}
For $m_{\rm c}^{\star}=1.4{\rm GeV}$ and $\Lambda_{\overline{\rm
MS}}=300{\rm MeV}$ the NLO result is by about 20\% larger than the LO
one.
So the correction turns out to be quite sizable.
To estimate something like a ``theoretical error'', we look at
$\eta_1^{\star}$'s dependence on the three matching scales for fixed
values of $m_{\rm c}^{\star}$ and $\Lambda_{\overline{\rm MS}}$.
The variation with $\mu_{\rm c}$ appears to be the largest one, it is
plotted in \fig{fig:scaledep}.
It is clearly seen, that the scale dependence in NLO is about 50\%
smaller than in LO.
E.g.\ for the reference values of $\Lambda_{\overline{\rm MS}}=0.3{\rm
GeV}$, $m_{\rm c}^{\star}=1.4{\rm GeV}$, $\mu_{\rm b}=4.8{\rm GeV}$,
$\mu_{\rm W}=\MW=80{\rm GeV}$ the variation of $\mu_{\rm c}$ from 1.1
to 1.7 GeV amounts to a change of 63\% in the LO case and 34\% in the
NLO result.
\begin{eqnarray}
	\eta_{1}^{\star} &=& 1.01
	\begin{array}{l}
	\scriptstyle+0.42\\[-2.1truemm]\scriptstyle-0.21
	\end{array}
	\mbox{(LO)}
\nn \\
	\eta_{1}^{\star} &=& 1.19
	\begin{array}{l}
	\scriptstyle+0.23\\[-2.1truemm]\scriptstyle-0.17
	\end{array}
	\mbox{(NLO)}
\end{eqnarray}
Fixing $\mu_{\rm c}=m_{\rm c}^{\star}=1.4{\rm GeV}$, we further
analyze the dependence of $\eta_{1}^{\star}$ on the scale $\mu_{\rm
W}$.
Varying $\mu_{\rm W}$ from 60 to 100 GeV, we find a change of
$\eta_{1}^{\star}$ by 7\% in LO and 4\% in NLO.
This residual scale dependence is therefore much weaker than the one
for $\mu_{\rm c}$.
The residual dependence on the matching scale $\mu_{\rm b}$ turns out
to be completely negligible, the extreme choice $\mu_{\rm b}=\mu_{\rm
W}=\MW$, which means neglecting completely the effects from an
effective five flavour theory, leads to an error of the order of 1\%.

We now want to discuss the implications of $\eta_{1}^{\star}$ on the
short distance contribution to the ${\rm K}_{\rm L}$-${\rm K}_{\rm S}$
mass difference.
To a very good approximation the part stemming only from the first two
generations is given by
\cite{BurasSlominskiSteger:84}
\begin{eqnarray}
	\Delta m_{\rm K}^{\rm SD,c}
	&=&
	\frac{\GF^2}{6\pi^2}m_{\rm K} f_{\rm K}^2 B_{\rm K}
	\left({\rm Re} \lambda_{\rm c}\right)^2 m_{\rm c}^{\star 2}
	\eta_{1}^{\star}
	.
\label{MassDiff}
\end{eqnarray}
For the input parameters $m_{\rm K}=0.498{\rm GeV}$, $f_{\rm
K}=0.161{\rm GeV}$, ${\rm Re}\lambda_{\rm c}=0.215$, $\GF=1.167\times
10^{-5}{\rm GeV}^{-2}$ and $B_{\rm K}=0.7$ the mass difference $\Delta
m_{\rm K}^{\rm SD,c}$ is given in \tab{tab:deltaM}.
Since $\Delta m_{\rm K}^{\rm SD}$ depends linearly upon the
nonperturbative parameter $B_{\rm K}$ the result may be easily
rescaled to other values of this parameter.
Note, that if we take $\Lambda_{\overline{\rm MS}}=0.3{\rm GeV}$ and
$m_{\rm c}^{\star}=1.4{\rm GeV}$, the short distance contribution of
the charm sector to the total mass difference is as much as 64\%.
The terms containing the top quark contribute another 6\%, therefore
the short distance physics is able to reproduce about 70\% of the
observed mass difference, which is much more than previously thought.



\end{document}